\documentstyle[aps,floats]{revtex}
\input{psfig.tex}
\begin{document}
\draft
\preprint{\vbox{\hbox{RAP-235} 
                \hbox{astro-ph/9805173}
}}

\title{Efficient Computation of Hyperspherical Bessel Functions}

\author{Arthur Kosowsky\footnote{kosowsky@physics.rutgers.edu}}
\address{Department of Physics and Astronomy, Rutgers University,
136 Frelinghuysen Road, Piscataway, New Jersey~~08854-8019}
\date{May 13, 1998}
\maketitle

\begin{abstract}
Fast and accurate computations of the power spectrum of
cosmic microwave background fluctuations are essential for
comparing current and upcoming data sets with the large
parameter space of viable cosmological models. The most efficient
numerical algorithm for power spectrum calculation, recently
implemented by Seljak and Zaldarriaga, involves integrating
sources against spherical Bessel functions or, in the cases
of a non-flat universe, analogous hyperspherical Bessel functions.
Evaluation of these special functions usually dominates the
computation time in non-flat spatial geometries. This paper presents
a highly accurate and very fast WKB approximation for 
computing hyperspherical
Bessel functions which will greatly increase the speed of
microwave background power spectrum computations in open and
closed universes.
\end{abstract}

\pacs{02.30.Gp, 02.30.Hq, 98.70.Vc \hspace*{1 cm} RAP-235, 
astro-ph/9805173}

\section{INTRODUCTION}

The power spectra of temperature and polarization
fluctuations of the cosmic microwave background contain a rich harvest of
cosmological information. From upcoming high-resolution maps of
the microwave background, we can hope to determine the basic
cosmological parameters to high precision, as well as test the
nature of the primordial perturbations and the mechanism for
structure formation in the Universe \cite{parameters}. 
Extracting this information
will require computationally intensive analysis of the parameter
space of cosmological models; a basic, general family of inflation-like
models requires around ten parameters. Monte Carlo explorations of
such a large parameter space will require power spectra evaluations
for millions of models, so it is imperative that the fastest
possible code be available for calculating power spectra.

Traditionally, microwave background power spectra were computed
by expanding the angular dependence of the radiation field in
multipole moments and then evolving the resulting set of coupled
differential equations. For power spectra at sub-degree scales,
typically several thousand coupled equations need to be evolved;
the CPU time for such codes on current workstations is generally 
measured in hours. Even scaling up to supercomputer speeds, large
Monte Carlo calculations are not possible with such codes.

A major advance was the realization by Seljak and Zaldarriaga that
a formal integral solution to the photon evolution equation allows
the solution (in flat space) to be expressed as a source term 
integrated against
spherical Bessel functions \cite{lineofsight}. 
The source term still needs to be
calculated via coupled evolution equations, but since only the
lowest moments contribute to the source term, far fewer equations
must be evolved. In essence, the full set of coupled equations
integrates the Bessel equation, whose solution is already known. 
The spherical Bessel functions depend on radius $r$ and on two
parameters, the wavenumber $k$ and the index $l$; however, in
flat space, the wavenumber can be scaled with the radial variable
so $j_l(kr)$ is actually a one-parameter family of functions.
Microwave background computations generally require values of $l$
up to a few thousand, and it is computationally feasible to 
precompute and cache the required Bessel functions.
Thus the flat-space power spectrum can be evaluated with a greatly
increased speed using the Seljak-Zaldarriaga algorithm.

For open or closed universes, the analogous ``hyperspherical'' Bessel
functions depend on the radial coordinate and the wavenumber
separately, so an inordinate number of functions would need to
be precomputed and stored. Seljak and Zaldarriaga resort
to a numerical integration of the differential equation defining
the hyperspherical Bessel functions, which is far slower than 
calling the precomputed functions in the flat case. This paper calculates
the WKB approximation to the hyperspherical Bessel functions, which,
while not as fast as precomputation, offers a dramatic increase
in computational speed over other methods. Furthermore, the WKB
approximation is {\it highly} accurate for all but the lowest
few values of $l$.

The following Section gives a brief review of WKB theory. Section
III then gives an overview of the relevant properties of the
hyperspherical Bessel functions; Section IV derives the WKB approximation
for all cases, including the usual spherical Bessel functions in
the flat space limit. Comparisons with exact functions
demonstrate the remarkable accuracy of the approximation. 
The paper concludes with brief remarks discussing implementation
and application of the approximations.

\section{REVIEW OF WKB THEORY}

The WKB approximation applies to equations of the Schrodinger form
\begin{equation}
\epsilon^2 y'' = Q(x) y.
\label{schrodinger}
\end{equation}
Such equations have exponential behavior in regions where $Q(x)>0$
(dissipative regions) and oscillatory behavior when $Q(x) < 0$
(dispersive regions).  Points with $Q(x)=0$ are called turning points,
marking the transition between the two behaviors. The WKB approximation
(away from the turning points) consists of an exponentiated power series
in $\epsilon$:
\begin{equation}
y(x) = \exp\left[{1\over \epsilon}\sum_{n=0}^\infty 
                             S_n(x) \epsilon^n \right].
\label{wkbdef}
\end{equation}
This paper considers the usual first-order WKB approximation, which
retains only the $n=0$ and $n=1$ terms of the expansion. This
approximation is often surprisingly accurate for $\epsilon$
as large as 1.
For a more detailed exposition of WKB theory, see Ref.~\cite{bender}.

Equations possessing a single turning point are the most straightforward
application of WKB theory. Without loss of generality, translate
the dependent variable so that the turning point is at the origin
and $Q(x) > 0$ for $x>0$. In the dissipative region $x>0$, which
will be called region I, a simple
asymptotic expansion gives the familiar ``WKB formula'' for the
exponentially decaying solution
\begin{equation}
y_I(x) \simeq C\, \left[Q(x)\right]^{-1/4}\exp\left[ -{1\over\epsilon}
                          \int_0^x \sqrt{Q(t)}\, dt\right],
\label{wkbI}
\end{equation}
where $C$ is an undetermined normalization constant. In
region II in the neighborhood of $x=0$, $Q(x)$ can be replaced by
its asymptotic expansion $Q\sim ax,$ $x\rightarrow 0$ which converts
Eq.~(\ref{schrodinger}) to an Airy equation with solution
\begin{equation}
y_{I\!I}(x)\sim 2C\sqrt{\pi} (a\epsilon)^{-1/6} 
     {\rm Ai}\left(a^{1/3}\epsilon^{-2/3}x\right),\qquad x\rightarrow 0
\label{wkbII}
\end{equation}
where ${\rm Ai}(x)$ is the first type of Airy function and
the prefactor has been determined by an asymptotic match with
Eq.~(\ref{wkbI}). Finally, in region III with $x < 0$, an analytic
continuation of Eq.~(\ref{wkbI}) and an asymptotic match to $y_{I\!I}$
gives
\begin{equation}
y_{I\!I\!I}(x) \simeq 2C \left[Q(x)\right]^{-1/4}
        \sin\left[ {1\over\epsilon}
        \int_x^0 \sqrt{-Q(t)}\, dt + {\pi\over 4}\right].
\label{wkbIII}
\end{equation}
Remarkably, there exists a single function which uniformly
approximates $y(x)$ and reduces to the above asymptotic forms in
the various limits:
\begin{eqnarray}
y(x) &\simeq& 2\sqrt{\pi} C \left({3S_0(x)\over 2\epsilon}\right)^{1/6}
\left[Q(x)\right]^{-1/4} {\rm Ai}\left[
\left({3S_0(x)\over 2\epsilon}\right)^{2/3}\right],\\
\label{langer}
&&\qquad\qquad S_0(x) = \int_0^x \sqrt{Q(t)}\, dt.\nonumber
\end{eqnarray}
This formula is the basis for all approximations presented
in this paper.

For the case of a function with two turning points 
at $x=A$ and $x=B$, two
single-turning-point solutions can be asymptotically
matched to form the general solution, which
leads to the eigenvalue condition
\begin{equation}
{1\over\epsilon}\int_A^B \sqrt{-Q(t)}\, dt = \left(n+{1\over 2}\right)\pi
\label{eigencondition}
\end{equation}
with $n$ a non-negative integer. This condition
typically is used to approximate
the energy eigenvalues of a particle in an arbitrary potential well
in quantum mechanics.

\section{PROPERTIES OF HYPERSPHERICAL BESSEL FUNCTIONS}

It is straightforward to apply these expressions to the case of
hyperspherical Bessel functions. A review of their relevant
properties is presented here; a more detailed exposition of these
functions on which the following discussion is based 
is given in Ref.~\cite{abbot}. The hyperspherical
Bessel functions $\Phi_l^\beta(r)$
are radial eigenfunctions of the covariant Laplace
operator in spherical coordinates, for spaces of constant
curvature:
\begin{equation}
(\nabla^2 + k^2) \Phi_l^\beta(r) Y_{lm}(\theta,\phi) = 0
\label{laplace}
\end{equation}
where the index $\beta = \sqrt{k^2 + K}$
with $K=H_0^2(\Omega_0 - 1)$ the spatial curvature, and 
$\nabla$ is the covariant 3-derivative associated with
the spatial part of the Robertson-Walker metric:
\begin{equation}
ds^2 = dt^2 - R^2(t)\left[{dr^2\over 1-Kr^2} + r^2 d\Omega^2\right].
\label{metric}
\end{equation} 
In this paper, all physical
distances will be expressed in units of the curvature scale,
giving $K=1$ for a closed universe, $K=0$ for a critical (flat)
universe, and $K= -1$ for an open universe. 
The eigenfunctions $\Phi_l^\beta$, termed 
``hyperspherical'' or ``ultraspherical'' Bessel functions
because they reduce to the usual spherical Bessel functions in the
case of a flat universe, are very useful because they
serve as the analog of Fourier modes for cosmological
models with non-zero curvature. For a critical density universe
with no spatial curvature, $\nabla$ is just the usual gradient
operator and $\Phi_l^\beta(r) = j_l(\beta r) = j_l(kr)$. 

It is convenient to change variables to the radial coordinate $\chi$
defined by $d\chi = dr/\sqrt{1-Kr^2}$; explicitly,
\begin{equation}
r(\chi) = \cases{\sin\chi,& $K=1$;\cr
              \chi,& $K=0$;\cr
              \sinh\chi,& $K= -1$.\cr}
\label{chidef}
\end{equation}
Then it is straightforward to demonstrate from Eq.~(\ref{laplace})
that the hyperspherical Bessel functions satisfy the Schrodinger
equation
\begin{equation}
{d^2 u_l^\beta\over d\chi^2} = 
\left[{l(l+1)\over r(\chi)^2} - \beta^2\right] u_l^\beta
\label{ueq}
\end{equation}
where $u_l^\beta(\chi) = r(\chi)\Phi_l^\beta(\chi)$. 
The hyperspherical Bessel functions are normalized so as
to match the normalization of $j_l(kr)$ in the flat-space limit.
For the cases $K=0$ and $K= -1$, the momentum variable $\beta$ can have
any positive value, Eq.~(\ref{ueq}) has a single turning point
for $\chi > 0$, and the WKB approximation can be applied directly.

The $K=1$ case is more complicated: the spatial sections of the
spacetime are compact, resulting in a discrete eigenvalue spectrum
for the eigenfunctions. It is possible to express the closed-universe
functions in terms of associated Legendre polynomials as
\begin{equation}
\Phi_l^\beta \propto (\sin\chi)^{-1/2} P^{-l-1/2}_{\beta - 1/2}(\cos\chi);
\label{pmnform}
\end{equation}
this form shows that $\Phi_l^\beta$ is periodic in $\chi$ with period $2\pi$.
As with spherical Bessel functions, $\Phi_l^\beta$ is symmetric 
(antisymmetric) around $\chi=0$ for $l$ even (odd); 
thus the function is determined by
its values on the interval $[0,\pi]$. Requiring $\Phi_l^\beta$ to be
single-valued gives the condition
\begin{equation}
\Phi_l^\beta(-\cos\chi) = \cos[(\beta-l-1)\pi]\Phi_l^\beta(\cos\chi).
\label{bdrycond}
\end{equation}
Two conclusions follow immediately: $\beta$ must be a positive integer,
and the functions are symmetric (antisymmetric) around $\chi=\pi/2$
if $\beta-l-1$ is even (odd). Thus the value of $\Phi_l^\beta(\chi)$
must be computed only on the interval $[0,\pi/2]$ and can be determined
for all other values of $\chi$ by symmetry.
It can be demonstrated that $\beta=1$ and $\beta=2$ represent gauge
modes, not physical perturbations \cite{lifshitz}, so $\beta$ takes
integer values of 3 or larger, and $\beta > l$ follows from
Eq.~(\ref{pmnform}).

The hyperspherical Bessel functions also satisfy several useful
recursion relations which are given in Ref.~\cite{abbot}. A closed-form
expression for $\Phi_l^\beta$ can be obtained as $l+1$ derivatives
of an elementary function; it is practicable to use these exact
expressions for evaluating the hyperspherical Bessel functions
up to $l=4$ or 5. For precise calculation of the functions for
larger values of $l$, recursion techniques can be used for the
open case, in analogy with the Miller's method evaluation of
Bessel functions in Ref.~\cite{numrec}. However, for the
closed case, downwards recursion is not always available due to
the restriction $l < \beta$, so direct integration of
Eq.~(\ref{ueq}) is better \cite{openlineofsight}.

\section{WKB APPROXIMATIONS}

To apply the WKB formalism most effectively, first divide both sides
of Eq.~(\ref{ueq}) by $l(l+1)$ to obtain Eq.~(\ref{schrodinger})
with
\begin{equation}
Q(\chi) = {1\over r^2(\chi)} - \alpha^2
\label{Qdef}
\end{equation}
with $\alpha\equiv \beta\epsilon$ and $\epsilon = 1/\sqrt{l(l+1)}$.
Turning points $\chi_0$ are located where 
\begin{equation}
r(\chi_0) = 1/\alpha; 
\label{chi0def}
\end{equation}
note that the $K=0$ and $K=1$ cases each have a single turning point,
while the $K=1$ case has a single turning point within the range
$[0,\pi/2]$. Thus Eq.~(\ref{langer}) can be applied directly
to each case. The
WKB approximation is an asymptotic series in powers of $\epsilon$,
which in this case is the inverse of $l$. For larger $l$ values,
the approximation becomes progressively better; it is also better
away from the region of the turning point where the various asymptotic
solutions are matched. As demonstrated below, the approximation is
remarkably good even for $l=2$ and 3.
 
The WKB approximation offers a great increase in numerical speed
because the required integrals can be performed exactly in terms
of elementary functions. For the $K=0$ case,
\begin{mathletters}
\begin{eqnarray}
\int_{\chi_0}^\chi \left(\alpha^2 - {1\over t^2}\right)^{1/2} dt
   &=& \sqrt{\alpha^2\chi^2 - 1} - \sec^{-1}(\alpha\chi),\\
\int_{\chi}^{\chi_0} \left({1\over t^2} - \alpha^2\right)^{1/2} dt
   &=& \log\left({1+\sqrt{1-\alpha^2\chi^2}\over\alpha\chi}\right)
        - \sqrt{1-\alpha^2\chi^2}.
\end{eqnarray}
\label{flatQints}
\end{mathletters}
For the $K= -1$ case, defining $w \equiv \alpha\sinh\chi$,
\begin{mathletters}
\begin{eqnarray}
\int_{\chi_0}^\chi \left(\alpha^2 - {1\over \sinh^2 t}\right)^{1/2} dt
  &=& \alpha\log\left({\sqrt{w^2-1} 
            + \sqrt{w^2+\alpha^2}\over\sqrt{1+\alpha^2}}\right)
      + \tan^{-1}\left[{1\over\alpha}
            \left({w^2+\alpha^2\over w^2-1}\right)^{1/2}\right] 
      - {\pi\over 2},\\
\int_{\chi}^{\chi_0} \left({1\over \sinh^2 t} - \alpha^2\right)^{1/2} dt
  &=& {\alpha\over 2}\tan^{-1}
      \left({2\sqrt{(1-w^2)(w^2+\alpha^2)}\over 2w^2+\alpha^2-1}\right)
      + \log\left({\alpha\sqrt{1-w^2}+\sqrt{\alpha^2+w^2}\over 
                         w^2\sqrt{1+\alpha^2}}\right).
\end{eqnarray}
\label{openQints}
\end{mathletters}
And similarly, for the $K=1$ case, defining $v\equiv\alpha\sin\chi$,
\begin{mathletters}
\begin{eqnarray}
\int_{\chi_0}^\chi \left(\alpha^2 - {1\over \sin^2 t}\right)^{1/2} dt
  &=& \tan^{-1}\left[{1\over\alpha}
            \left({\alpha^2-v^2\over v^2-1}\right)^{1/2}\right] 
         -{\alpha\over 2}\tan^{-1}\left({2\sqrt{(v^2-1)(\alpha^2-v^2)} 
            \over 2v^2-\alpha^2-1}\right)
      - {\pi\over 2},\\
\int_{\chi}^{\chi_0} \left({1\over \sin^2 t} - \alpha^2\right)^{1/2} dt
  &=& \tanh^{-1}\left[\alpha
      \left({1-v^2\over \alpha^2 -v^2}\right)^{1/2}\right]
      - \log\left({\sqrt{1-v^2}+\sqrt{\alpha^2-v^2}\over 
                         \sqrt{\alpha^2 -1}}\right).
\end{eqnarray}
\label{closedQints}
\end{mathletters}
Care must be taken to use the correct branch of the inverse tangent
functions.

In the closed case, the eigenvalue condition
Eq.~(\ref{eigencondition}) becomes
\begin{equation}
\beta = n+{1\over 2} + {1\over \epsilon}
\label{betacorr}
\end{equation}
with $n$ an integer, so for a given exact integer value of $\beta$,
the corrected WKB eigenvalue can be obtained by the replacement
\begin{equation}
\beta\rightarrow \beta - {1\over 8l} + {1\over 16l^2},
\label{betareplace}
\end{equation}
which is sufficiently accurate for a first-order WKB approximation.
If the eigenvalue $\beta$ is not corrected when evaluating the 
approximate function, the function or its first derivative
will be discontinuous at $\chi=\pi/2$.

The other necessary numerical ingredient is the evaluation of
the Airy function in Eq.~(\ref{langer}). If a fast routine is not available,
a reasonable approximation is to use the leading asymptotic
behavior at large arguments and a Taylor series around the origin.
With a crossover at $|x|=1.6$ and a series including $x^{13}$ terms,
the residual error in ${\rm Ai}(x)$ 
is at the 1\% level. Fast routines
based on Chebyshev polynomial fits or Pade expansions are readily 
obtainable in both Fortran and C. Note that since the
Airy function is evaluated for every hyperspherical Bessel function,
errors in Airy function evaluation translate into systematic
errors in $\Phi^\beta_l$. Using the simple asymptotic approximation
to ${\rm Ai}(x)$, for example, leads to a systematic 1\% error
when integrating smooth functions against $\Phi^\beta_l$ independent
of $l$; this error is reduced to 0.01\% or less when an accurate Airy
function evaluation is employed \cite{huprivate}.

\begin{figure}[htbp]
\centerline{\psfig{file=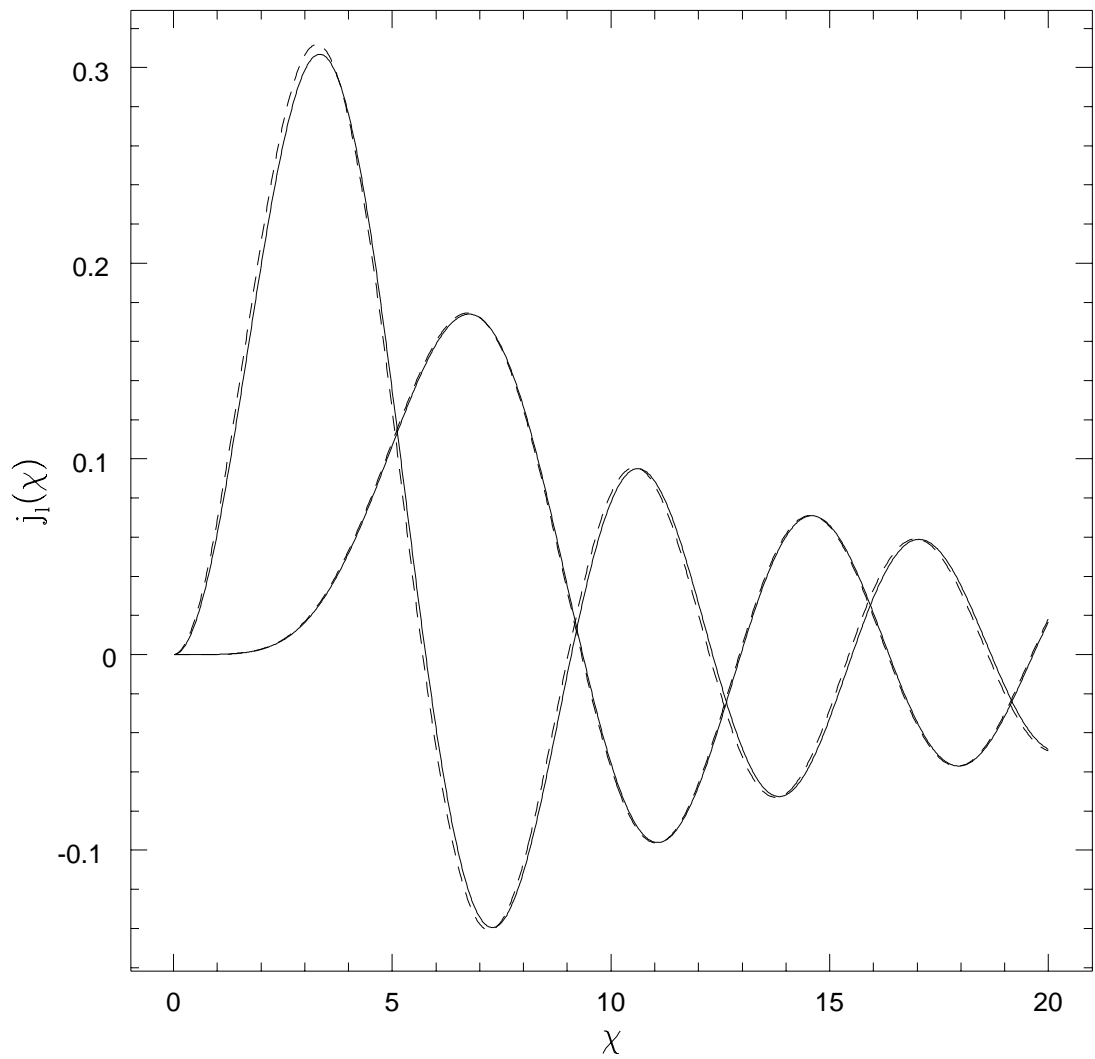,width=6in}}
\caption{The WKB approximation for the usual spherical
Bessel functions $j_l(\chi)$, for $l=2$ and $l=5$. The
exact functions are solid lines, the approximate
functions dashed lines. For $l>5$ 
the exact and approximate functions are indistinguishable
on the scale of the plot.
}
\label{flat1}
\end{figure}

Figure 1 displays the exact and WKB-approximated spherical Bessel
functions $j_l(\chi)$ for $l=2$ and $l=5$. The accuracy is very
good even for $l=2$, with an error of 1.5\% at the first peak
in the $l=2$ case and 0.6\% at the first peak in the $l=5$ case.
For higher values of $l$, the actual and approximated functions
are indistinguishable on the scale of the plot; by $l=20$, the
first peak is accurate to 0.05\%. As mentioned above, the harmonics
for the lowest few $l$ values
can be evaluated exactly in terms of trigonometric functions.

\begin{figure}[htbp]
\centerline{\psfig{file=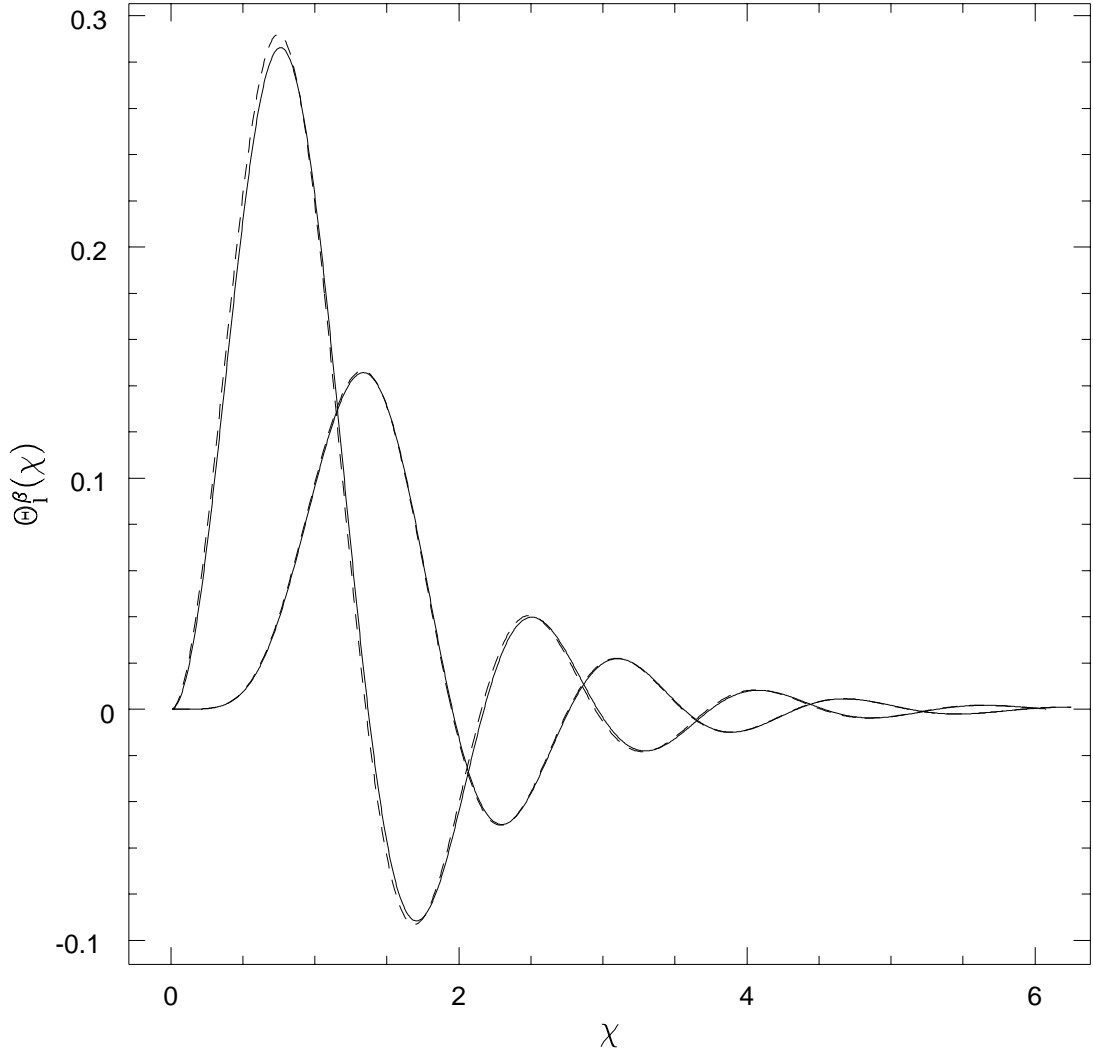,width=6in}}
\caption{The WKB approximation for the open universe
hyperspherical Bessel functions $\Phi^4_2$ and $\Phi^4_5$, shown 
as the dashed lines, compared with the solid exact functions.
}
\label{open1}
\end{figure}

\begin{figure}[htbp]
\centerline{\psfig{file=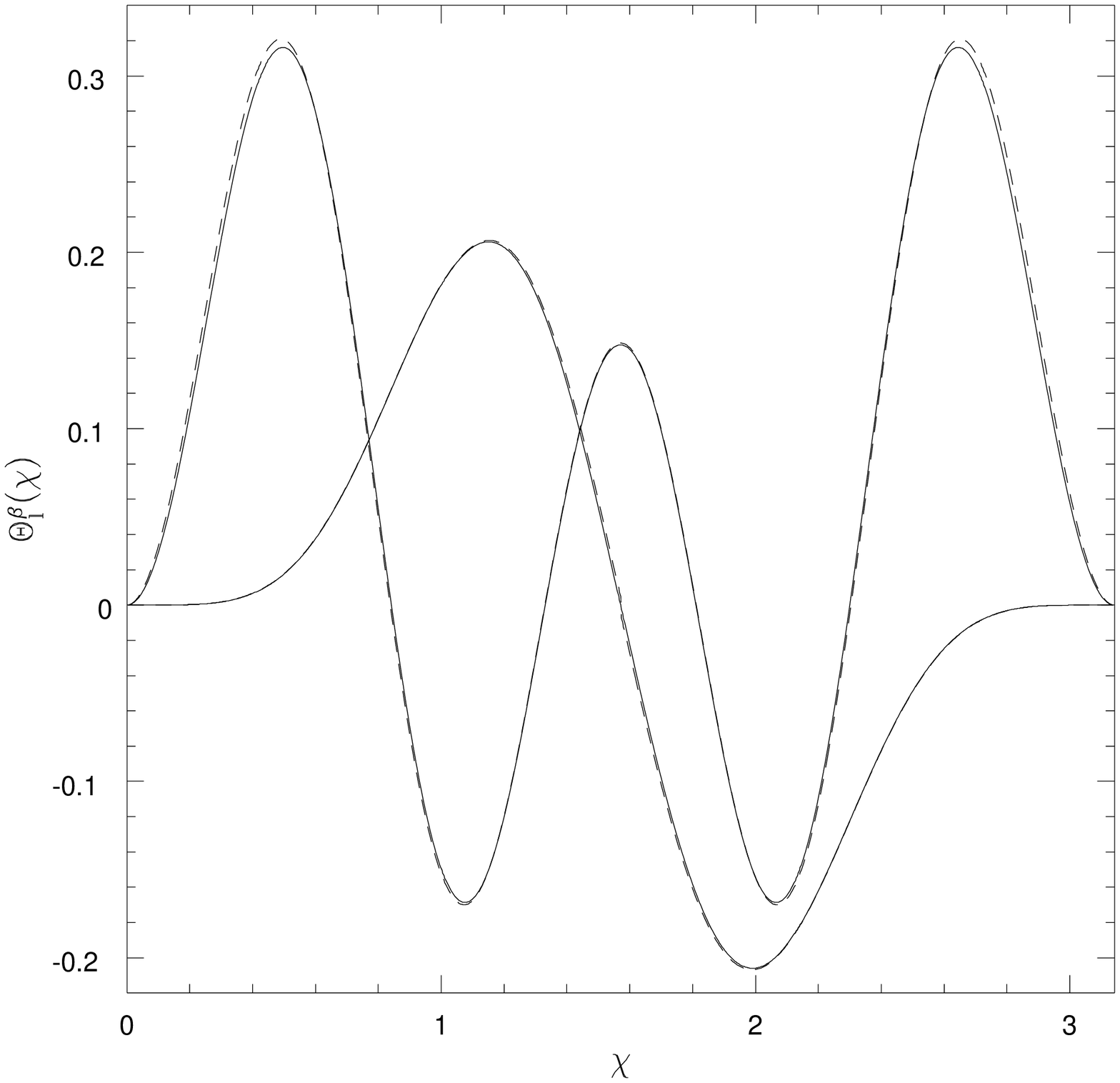,width=6in}}
\caption{
The WKB approximation for the closed universe
hyperspherical Bessel functions $\Phi^7_2$ and $\Phi^7_5$, shown 
as the dashed lines, compared with the solid exact functions.
}
\label{closed1}
\end{figure}

Figures 2 and 3 compare the exact and WKB-approximated hyperspherical
Bessel functions for open and closed universes, respectively.
The level of accuracy is essentially the same as in the flat
case, with approximations for $l>5$ indistinguishable from the
exact values on the scales of the plots. One particular set of
closed-universe
functions always has a significant error: the lowest eigenvalue
corresponding to a given $l$ with the unperturbed value $\beta=l+1$.
As shown in Figure 4, the approximate functions have discontinuous
derivatives at $\chi=\pi/2$ for all $l$. The reason is that
the turning point $\chi_0$ is so close to $\pi/2$ for the lowest
eigenvalue that the solution does not have enough room between
$\chi_0$ and $\pi/2$ to attain
its asymptotic behavior away from the turning point, so an
asymptotic match around $\chi=\pi/2$ is unsuccessful. This behavior
can be circumvented with an exact form for the lowest eigenfunctions:
\begin{eqnarray}
\Phi_l^{l+1}(\chi) &=& \left[ (2l)!!\over (l+1)(2l+1)!!\right]^{1/2}
\sin^l\chi\qquad\qquad(K=1)\nonumber\\
&\sim & \left[2\pi\over 2l+1\right]^{1/4}
\left[l\over (l+1)(2l+1)\right]^{1/2}
\left[ 1 + {9\over 48l} - {7\over 512l^2} \right]\sin^l\chi,
\qquad\qquad l\rightarrow\infty;
\label{lowesteigenfcn}
\end{eqnarray}
the asymptotic expression is good to 0.02\% at $l=2$.
Fortunately,
the second and higher eigenvalues are not greatly affected
by this problem (see Fig.~3, for example).

\begin{figure}[htbp]
\centerline{\psfig{file=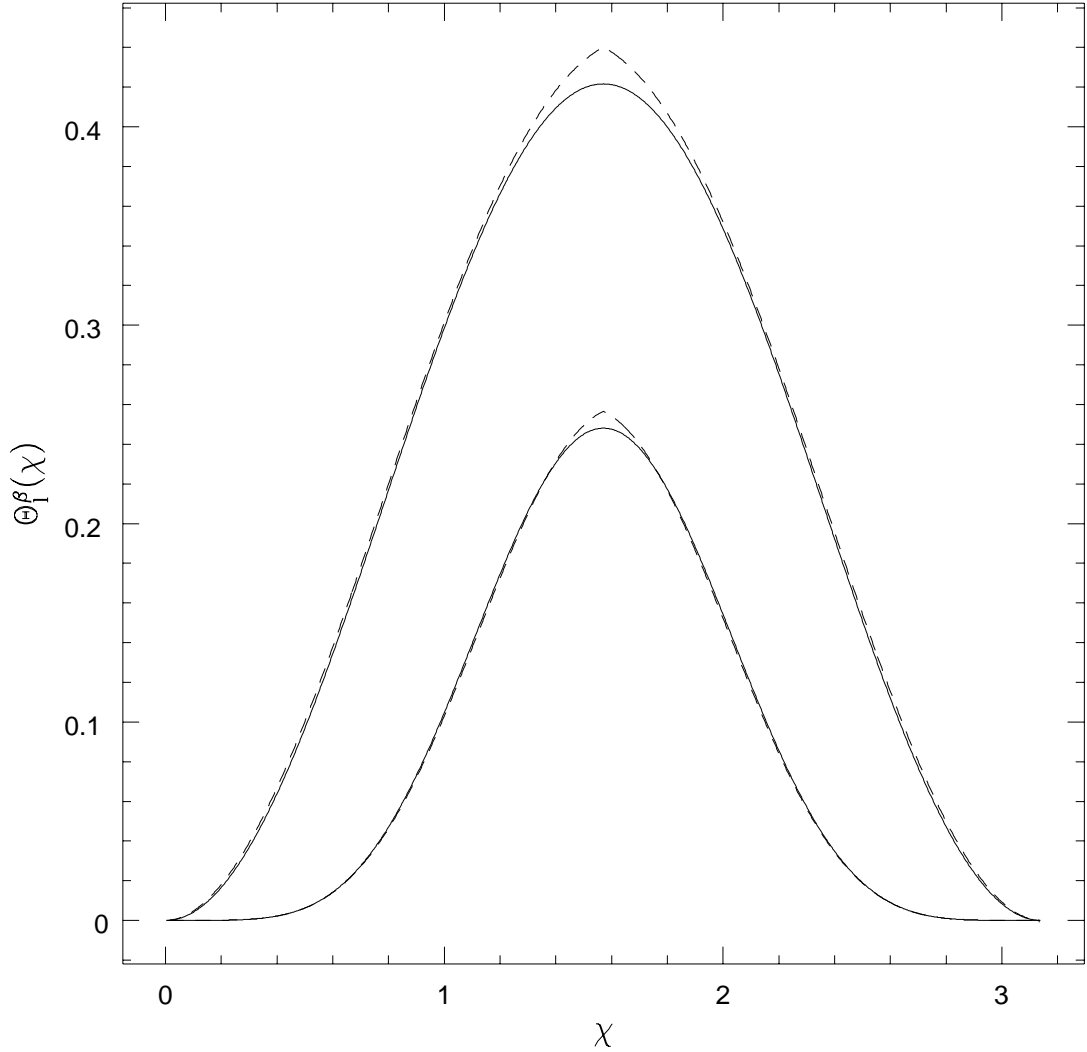,width=6in}}
\caption{
The WKB approximation for the closed universe
hyperspherical Bessel functions $\Phi^3_2$ and $\Phi^6_5$, shown 
as the dashed lines, compared with the solid exact functions. The
discrepancy near $\chi=\pi/2$ is because the matching solution does
not have enough space between the two turning points to reach its
asymptotic value. This artifact persists in the lowest eigenfunction
for all values of $l$.
}
\label{closed2}
\end{figure}

\section{CONCLUDING REMARKS}

The above calculations demonstrate that the WKB approximation to
hyperspherical Bessel functions is highly accurate, with increasing
accuracy as $l$ becomes larger. Furthermore, the computation of
one of the functions at a given value of $\chi$ requires around
ten elementary function calls, plus some arithmetic, so the
approximation is much faster than evaluation based on recursion
methods or integration of Eq.~(\ref{ueq}). The computation time is
also independent of $l$, in marked contrast to other methods, and
is completely stable for any values of $l$ and $\beta$. This
approximation will substantially speed the computation of
microwave background power spectra in open or closed universes,
potentially by an order of magnitude based on rough timings
of the CMBFAST code \cite{cmbfast}.

A further speedup in evaluating these functions can be accomplished
by precomputing and caching the integrals in Eqs.~(\ref{flatQints})
to (\ref{closedQints}); each is a function of the two variables $\chi$
and $\alpha$. The integrals are very uneventful functions of these
variables, and accurate interpolation is possible based on a small
number of computed values. With this scheme, the computational work
for a hyperspherical Bessel function is reduced to evaluating a power
and a sine or exponential in Eq.~(\ref{wkbI}) or (\ref{wkbIII}), 
an inverse sine or hyperbolic sine to evaluate $\chi_0$,
and a small amount of arithmetic.

The uniform WKB approximation presented here is useful for evaluating
any functions defined by a second-order differential equation in
Schrodinger form, especially when speed of evaluation is more
important than accuracy to many significant figures. 
This is often the case in physics
problems, and particularly in the case where a special function
is integrated against another function. If it is impracticable
to cache all needed values of the special function, WKB offers
a fast, simple, and remarkably accurate alternative.


\end{document}